\documentclass[
reprint,
superscriptaddress,
 amsmath,amssymb,
 aps,
 nofootinbib
]{revtex4-2}

\usepackage{graphicx}% Include figure files
\usepackage{dcolumn}% Align table columns on decimal point
\usepackage{bm}% bold math
\usepackage{xcolor}
\usepackage[normalem]{ulem}

\def\l{\left(}
\def\r{\right)}
\def\lz{\left[}
\def\rz{\right]}

\begin{document}

\preprint{APS/123-QED}

\title{Next-to-leading-order solution to Kerr-Newman black hole superradiance}

\author{Shou-Shan Bao}
\email{ssbao@sdu.edu.cn}
\affiliation{Institute of Frontier and Interdisciplinary Science,
Key Laboratory of Particle Physics and Particle Irradiation (MOE),
Shandong University, QingDao 266237, China}

\author{Qi-Xuan Xu}
\email{qixuan.xu22@imperial.ac.uk}
\affiliation{Theoretical Physics, Blackett Laboratory, Imperial College, London SW7 2AZ, United Kingdom}

\author{Hong Zhang}
\email{hong.zhang@sdu.edu.cn}
\affiliation{Institute of Frontier and Interdisciplinary Science,
Key Laboratory of Particle Physics and Particle Irradiation (MOE),
Shandong University, QingDao 266237, China}

\date{\today}

\begin{abstract}
The superradiant instabilities of Kerr-Newman black holes with charged or uncharged massive spin-0 fields are calculated analytically to the next-to-leading order in the limit of $\alpha\sim r_g \mu \ll 1$. A missing factor of $1/2$ in the previous leading-order result is identified. The next-to-leading order result has a compact form and is in good agreement with existing numerical calculations. The percentage error increases with $\alpha$, from a few percent for $\alpha\sim 0.1$ to about $50\%$ for $\alpha\sim 0.4$. Massive neutral scalars too heavy to be produced with Kerr black hole superradiance may exist in the superradiant region of Kerr-Newman black holes.
\end{abstract}

\maketitle

\section{\label{sec:Intro}Introduction}

Ultralight boson condensate could form around a rotating black hole (BH) if the boson's Compton wavelength is comparable to the size of the BH horizon. With the proper choice of parameters, such scalar condensate can continuously extract energy and angular momentum from the BH until the BH spin is below some critical value and/or nonlinear effects become important\cite{Arvanitaki:2010sy,Yoshino:2012kn,Baryakhtar:2020gao}. This phenomenon is known as BH superradiance \cite{Penrose:1969pc,Misner:1972kx}. There exist numerous works on various bosons, including spin-0 \cite{Detweiler:1980uk,Zouros:1979iw,Cardoso:2005vk,Konoplya:2006br,Dolan:2007mj,Konoplya:2011qq,Arvanitaki:2009fg,Arvanitaki:2010sy,Arvanitaki:2014wva,Arvanitaki:2016qwi,Yoshino:2013ofa,Yoshino:2015nsa,Brito:2014wla,Brito:2017zvb,Brito:2017wnc,Ficarra:2018rfu,Bao:2022hew,Roy:2021uye,Chen:2022nbb,Hui:2022sri,Witek:2012tr,Endlich:2016jgc}, spin-1 \cite{Baryakhtar:2017ngi,Dolan:2018dqv,East:2017mrj,East:2017ovw,East:2018glu,Endlich:2016jgc,Frolov:2018ezx,Pani:2012vp,Pani:2012bp,Siemonsen:2019ebd,Witek:2012tr,Percival:2020skc,Caputo:2021efm,East:2022ppo} and spin-2 \cite{Brito:2013wya,Brito:2020lup} fields. In this work, we focus on the ultralight scalars. The superradiance of other types of bosons could be found in the comprehensive review~\cite{Brito:2015oca}.

The scalar superradiance, especially with a Kerr BH, is important in phenomenology. Such BH-condensate systems have been widely studied for constraining the scalar properties and for the possible observation of the (gravitational wave) GW emission. It has been shown that the BH evolves along the Regge trajectories on the mass-spin plot if the superradiant effect is strong \cite{Arvanitaki:2010sy,Brito:2014wla}. Consequently, there are ``holes" on the Regge plot in which BHs cannot reside. Combing with the observed BH spin distribution, favored and unfavored scalar mass ranges can be identified \cite{Ng:2019jsx,Ng:2020ruv,Cheng:2022ula}. On the other hand, with the continuous GW generated by the BH-condensate, works have been done to study the possibility of resolving these systems from the backgrounds \cite{Arvanitaki:2010sy,Yoshino:2013ofa,Yoshino:2014wwa,Arvanitaki:2014wva,Arvanitaki:2016qwi,Brito:2017wnc,Brito:2017zvb}. The positive frequency drift \cite{Arvanitaki:2014wva,Baryakhtar:2017ngi} and the beatlike pattern \cite{Guo:2022mpr} have been proposed to distinguish them from other monochromatic GW sources, such as neutron stars. The unresolved BH-condensate systems have also been carefully studied as stochastic backgrounds for GW detectors \cite{Brito:2017wnc,Brito:2017zvb}.

The phenomenological study of BH superradiance depends on the accurate determination of the bound state's eigenfrequency. For Kerr BHs, the numerical continued fraction method was first proposed by Leaver for massless scalars \cite{Leaver:1985ax}. It is later developed for massive scalars in Ref.~\cite{Cardoso:2005vk} and further refined in Ref.~\cite{Dolan:2007mj}. In the small $\alpha\sim r_g\mu$ limit, an analytic approximation was obtained by Detweiler \cite{Detweiler:1980uk}. Nonetheless, these two solutions are not consistent with each other. The problem is recently resolved in our previous work by including the next-to-leading-order (NLO) contribution to the analytic approximation \cite{Bao:2022hew}. A power-counting strategy is also proposed which facilitates the NLO calculation.

In Ref.~\cite{Damour:1976kh}, Damour et.al. have shown that the superradiance can also be realized with a charged massive scalar field in Kerr-Newman spacetime. Comparably, it does not attract as much attention as that for Kerr BHs. It may be because the Kerr-Newman BH (KNBH) is unlikely to play important roles in astrophysics \cite{Gibbons:1975kk, Blandford:1977ds, Barausse:2014tra}. Nonetheless, as pointed out in Ref.~\cite{Zilhao:2014wqa}, the KNBH provides an ideal testing ground for studying the interplay between gravity and electrodynamics. In the previous studies of scalar superradiance with KNBHs, Detweiler's method has been applied to obtain the leading-order (LO) analytic approximation at the $\alpha\ll 1$ limit \cite{Furuhashi:2004jk,Hod:2014baa, Hod:2016bas}. The numerical solution has also been achieved using the 3-term continued fraction method \cite{Huang:2018qdl, Huang:2018qdl}. The parameter space of the KNBH superradiance is also probed by analyzing the existence of the potential well \cite{Myung:2022kex, Myung:2022gdb,Myung:2022biw}. 

In this work, we refine the power-counting strategy in our previous work and apply it to calculate the NLO contribution of the KNBH superradiance. A compact NLO expression for $\alpha\ll 1$ is obtained which could be straightforwardly applied to phenomenological study. The scalar field can be either neutral or charged. By comparing to the existing numerical results, the percentage error of the  NLO approximation increases with $\alpha$, from a few percent for $\alpha\sim 0.1$ to about $50\%$ for $\alpha\sim 0.4$. 
In comparison, the LO approximation does not agree with the numerical results qualitatively (see Fig.~\ref{fig:Comparison} below).

This paper is organized as follows. In Sec.~\ref{sec:KGeq}, we briefly review the Klein-Gordon equation to be solved and obtain the superradiance condition from its solution at the outer horizon. Detweiler's method is applied to derive the LO and NLO analytic expressions in Sec.~\ref{sec:AnalyticSolu}. In Sec.~\ref{sec:results}, the obtained analytic expressions are compared to the existing numerical calculation. Some effects relevant to phenomenology are also discussed. Finally, we summarize our results in Sec.~\ref{sec:Conclusion}.

\section{Scalars in Kerr-Newman Spacetime}\label{sec:KGeq}

The spacetime around a KNBH with mass $M$, angular momentum $J$ and charge $Q$ can be expressed in Boyer-Lindquist coordinates \cite{Boyer:1966qh},
%===
\begin{align}
\begin{aligned}
ds^{2} =& -\l 1-\frac{2r_gr-Q^{2}}{\Sigma^{2}}\r dt^{2} +\frac{\Sigma^{2}}{\Delta}dr^{2} + \Sigma^{2}d\theta^{2}\\
& + \lz (r^{2}+ a^{2}) + \frac{(2r_gr-Q^{2})a^{2}\sin^{2}\theta}{\Sigma^{2}}\rz \sin^{2}\theta d\varphi^{2}\\
& - \frac{2(2r_gr-Q^{2})a\, \sin^{2}\theta}{\Sigma^{2}}dtd\varphi,
\end{aligned}
\end{align}
%===
with
%===
\begin{subequations}
\begin{align}
a &= J/M,\\
r_{g} &= GM,\\
\Sigma^{2} &= r^{2}+a^{2}\cos^{2}\theta,\\
\Delta &= r^{2}- 2r_gr + a^{2} + Q^{2}.
\end{align}
\end{subequations}
%===
The equation $\Delta=0$ gives two event horizons at $r_\pm = r_g \pm b$ with $b=\sqrt{r_g^2-a^2-Q^2}$. In this work, we only consider the KNBHs with $r_g^2-a^2-Q^2\geq 0$.

To study the superradiance of a scalar field close to a BH, one needs to solve the combined Einstein and Klein-Gordon field equations, which is a very difficult task, especially because the existence of the scalar perturbs the spacetime around the BH. Nonetheless, it has been shown that this perturbation could be safely ignored due to the tiny energy-stress tensor of the scalar cloud for Kerr BH \cite{Brito:2014wla}. We assume the same situation happens for the KNBHs. We further assume the self-interaction of the scalar field can also be ignored. Then the problem reduces to solving the Klein-Gordon equation on the stationary Kerr-Newman background,
%===
\begin{align}
(\nabla^{\alpha}-iqA^{\alpha})(\nabla_{\alpha}-iqA_{\alpha})\phi - \mu^{2}\phi =0 \label{eq:KG-KN},
\end{align}
%===
where $\mu$ and $q$ are the mass and electric charge of the scalar field, respectively. The vector $A_\alpha$ is the background electromagnetic potential,
%===
\begin{align}
A_{\alpha} =  \frac{Qr}{\Sigma^{2}} \l -1, 0, 0, a\sin^{2}\theta \r.
\end{align}
%===
For complex scalars, $\phi$ can be written with the separation of variables,
%===
\begin{align}
\phi(t,r,\theta,\varphi) = \sum_{l,m}\int  d\omega R_{lm}(r)S_{lm}(\theta)e^{im\varphi}e^{-i\omega t}.\label{eq:SF-wavefunction}
\end{align}
%===
Inserting it into Eq.~\eqref{eq:KG-KN}, one obtains the angular equation,
%===
\begin{align}
\begin{aligned}
&\frac{1}{\sin\theta}\frac{d}{d\theta}\l \sin\theta\frac{dS_{lm}}{d\theta}\r + \\
&\lz-a^{2}(\mu^{2}-\omega^{2})\cos^{2}\theta -\frac{m^{2}}{\sin^{2}\theta} +\Lambda_{lm} \rz S_{lm} = 0\label{eq:Angular},
\end{aligned}
\end{align}
%===
where $\Lambda_{lm}$ is the eigenvalue. Its solution $S_{lm}(\theta)$ is called the spheroidal harmonic function, whose properties can be found in Ref.~\cite{Berti:2005gp}. The corresponding radial equation is~\cite{Press:1973zz},
%===
\begin{align}
\Delta \frac{d}{d r } \l\Delta \frac{d R_{lm}}{d r }\r + U(r)R_{lm} = 0,\label{eq:Teukolsky}
\end{align}
%===
with
%===
\begin{align}
\begin{aligned}
U(r) =& [\omega(a^{2}+r^{2})-am-qQr]^{2}\\
&+ \Delta[2am\omega -\mu^{2}r^{2}-a^{2}\omega^{2}-\Lambda_{lm}].
\end{aligned}
\end{align}
%===
These are the equations for the complex scalar field. For real scalars, one should set $q=0$ in Eq.~\eqref{eq:KG-KN} and choose only the real part on the right side of Eq.~\eqref{eq:SF-wavefunction}. In the rest of this paper, we focus on the equations for the complex scalars. The case for the real scalars can then be simply obtained by choosing $q=0$.

To obtain a constraint on the parameters that allow superradiance, we change to the tortoise coordinates,
%===
\begin{align}
dr_{*} = \frac{r^{2} + a^{2}}{\Delta}dr,
\end{align}
%===
with which the interesting region $r \in (r_{+}, + \infty)$ corresponds to $r_{*} \in (-\infty,+\infty)$. We also define,
%===
\begin{align}
R_{*}(r_{*}) = \sqrt{r^{2} + a^{2}}R(r).
\end{align}
%===
Then Eq.~\eqref{eq:Teukolsky} can be rewritten into a Schr\"{o}dinger-like equation,
%===
\begin{align}
\frac{d^{2}R_{*}(r_{*})}{dr^{2}_{*}} - V(r)R_{*}(r_{*}) = 0,\label{eq:SE*}
\end{align}
%===
where the effective potential is,
%===
\begin{align}
\begin{aligned}
V(r) =& - \left( \omega - \frac{am+qQr}{ a^{2}+ r^{2}} \right)^{2} 
+ \frac{\Delta\mu^{2}}{a^{2}+r^{2}}\\
& - \frac{\Delta}{(a^{2}+r^{2})^{2}} \lz 2am\omega - \Lambda_{lm} + a^{2}(\mu^{2}-\omega^{2}) \rz \\
&  + \frac{\Delta[\Delta + 2r(r-r_{g})]}{(a^{2}+r^{2})^{3}} - \frac{3\Delta^{2}r^{2}}{(a^{2}+r^{2})^{4}}.
\end{aligned}
\end{align}
%===
In the region close to the outer horizon $r_{+}$, the potential has the asymptotic form,
%===
\begin{align}
\lim_{r \to r_{+}}V(r)=-(\omega-\omega_{c})^{2}+\mathcal{O}(r-r_{+}),
\end{align}
%===
where the critical frequency is defined as
%===
\begin{align}
\omega_{c} = \frac{ma + qQr_{+}}{r_{+}^{2} + a^{2}}= \frac{ma + qQr_{+}}{2r_g r_{+}-Q^2}.\label{eq:CriticalFreq}
\end{align}
%===
Inserting this asymptotic expression of $V(r)$ into Eq.~\eqref{eq:SE*}, one gets the solution at the outer horizon,
%===
\begin{align}
\lim_{r_{*} \to - \infty} \!\!\! R_{*}(r_{*}) = d_{1}e^{-i(\omega-\omega_{c})r_{*}} + d_{2}e^{i(\omega-\omega_{c})r_{*}},
\end{align}
%===
where the first term is the wave falling into the outer horizon, and the second term is the wave escaping from the outer horizon; $d_1$ and $d_2$ are their respective amplitudes. Physically, nothing can escape from the horizon, indicating $d_2 = 0$. The superradiance requires the phase velocity and the group velocity to be in opposite directions, which leads to the superradiance condition for a KNBH,
%===
\begin{align}
\text{Re}(\omega)<\omega_{c}.\label{eq:SuperradianceCon}
\end{align}
%===
From Eq.~\eqref{eq:CriticalFreq}, we can see that with
$Q$ fixed, this condition is more relaxed (strict) compared to the superradiance condition of a Kerr BH if the charges of the scalar and the BH have the same sign (different signs).

\section{Analytic solution at $\alpha\ll 1$}\label{sec:AnalyticSolu}

In the small $\alpha$ limit, the asymptotic matching method first proposed in Ref.~\cite{Detweiler:1980uk} gives a reasonable approximation of the complex eigenfrequency $\omega$. In a previous work, we have further calculated the NLO contribution for Kerr BH superradiance~\cite{Bao:2022hew}. The NLO result has a much better agreement with the numerical solutions compared to the LO approximation. In the current work, we apply the method to KNBHs. In this section, we first repeat the LO approximation in Ref.~\cite{Huang:2018qdl}. A missing factor of $1/2$ is identified. We then continue to calculate the NLO contribution. The calculation is valid for both real and complex scalar fields. For a real scalar field, one simply sets $q=0$ throughout.

\subsection{Leading-order approximation}\label{sec:AnalyticSolu-LO}

We first formally introduce the power-counting parameter $\alpha \sim r_g \mu$ for the expansion. The scaling of other parameters are $\text{Re}\,\omega \sim \mu\sim q$ and $a\sim Q\sim r_+\sim r_- \sim r_g$. Unlike some previous calculations in which $\alpha$ is defined to be $r_g\mu$, here we leave $\alpha$ as a power-counting parameter, which could be $r_g\mu$ or any other quantity with the same scaling. In the limit $r \to +\infty $, the derivative term in Eq.~\eqref{eq:Teukolsky} divided by $\Delta^{2}$ can be written into a familiar form,
%===
\begin{align}
\begin{aligned}
\frac{1}{\Delta} \frac{d}{d r } \l\Delta \frac{d R}{d r }\r 
\approx \frac{d^{2} R}{d r^{2}} + \frac{2}{r}\frac{d R}{d r }= \frac{1}{r}\frac{d^{2}}{d r^{2}}(rR).
\end{aligned}
\end{align}
%===
The second term on the left side of Eq.~\eqref{eq:Teukolsky} divided by $\Delta^2$ can be expanded in powers of $r_{g}/r$. Keeping terms up to  $r_{g}^{2}/r^{2}$, the radial function at large $r$ limit $(r\gg r_{g})$ can be simplified as
%===
\begin{widetext}
\begin{align}
\begin{split}
\frac{d^{2}}{d r^{2}}(rR) + 
\lz(\omega^{2}-\mu^{2}) + \frac{2(2r_{g}\omega^{2}-r_{g}\mu^{2}-qQ\omega)}{r}
-\frac{l'(l'+1)}{r^{2}}+\mathcal{O}(r^{-3}) \rz rR = 0,\label{eq:SEwithCoulombPotential}
\end{split}
\end{align}
\end{widetext}
%===
where
%===
\begin{align}\label{eq:lp-def}
\begin{aligned}
l'(l'+1) =& \Lambda_{lm} + 4r_{g}^{2}(\mu^{2}-3\omega^{2})+ a^{2}(\omega^{2}-\mu^{2}) \\
&+ Q^{2}(2\omega^{2}-q^{2}-\mu^{2}) + 8r_{g}qQ\omega.
\end{aligned}
\end{align}
%===
The $l'$ is related to the orbital angular number by
%===
\begin{align}
    l'=l + \epsilon.
\end{align}
%===
Here $\epsilon\sim\mathcal{O}(\alpha^2)$ plays the role of a regulator and cannot be simply dropped.

For a confined profile, the real part of $\omega$ is less than the boson mass $\mu$. The physical solution is the one that decays exponentially at large $r$. It is more convenient to define,
%===
\begin{align}
\kappa &= \sqrt{\mu^{2}-\omega^{2}},\label{eq:kappa}\\
\lambda &= \frac{2r_{g}\omega^{2}-r_{g}\mu^{2}-qQ\omega}{\kappa}, \label{eq:lambda}\\
y &= \kappa r,\\
u(y) &= yR\left(\frac{y}{\kappa}\right).
\end{align}
%===
Then Eq.~\eqref{eq:SEwithCoulombPotential} can be rewritten as
%===
\begin{align}
\frac{d^{2}u(y)}{d y^{2}} + \lz-1 + \frac{2\lambda}{y} - \frac{l'(l'+1)}{y^{2}}\rz u(y) = 0.
\end{align}
%===
The two solutions are Whittaker functions, and only one of them has the correct behavior at $r\to +\infty$ required by the bound states. The solution with the correct behavior can be further written in terms of confluent hypergeometric functions. Finally, the radial function at large $r$ is
%===
\begin{align}
\begin{aligned}
R(r) = e^{-\kappa r}(2\kappa r)^{l'}U(l'+1-\lambda,2l'+2;2\kappa r),\label{eq:ana-R-large-r}
\end{aligned}
\end{align}
%===
up to an arbitrary normalization.

The bound states only exist if $\lambda > 0 $ in large $r$ region. The superradiance condition in Eq.~(16) gives another constraint $\omega(r_{+}^{2} + a^{2})-ma < qQr_{+}$. Combining these two inequalities, one can obtain,
%===
\begin{align}
\begin{aligned}
0 &<(2r_{g}\omega^{2}-r_{g}\mu^{2}-qQ\omega)r_+ \\
&<2r_{g}r_{+}\omega^{2}-r_{g}r_{+}\mu^{2}-(r_{+}^{2} + a^{2})\omega^{2}+ma\omega\\
&=ma\omega-r_{g}r_{+}\mu^{2}+Q^{2}\omega^{2}\\
&<\lz ma-(r_{g}r_{+}-Q^{2})\mu\rz\omega,
\end{aligned}
\end{align}
%===
indicating no superradiant bound state if $m \leq 0$. It also shows that Reissner-Nordstr\"om BHs could not hold bounded scalar clouds. The minimum KNBH spin $a$ allowing superradiant instability is approximately
$\mu(r_g r_{+}-Q^2)/m$.

Next, we look at Eq.~\eqref{eq:Teukolsky} in the small $r$ limit. For BH superradiance, the inner boundary is the outer horizon $r = r_{+}$. It is more convenient to write the radial function in terms of $z = (r-r_{+})/2b$, 
%===
\begin{align}
z(z+1)\frac{d}{dz}\left[ z(z+1)\frac{dR}{dz}\right]
+U(z) R = 0,
\end{align}
%===
where $U(z)$ can be written as an expansion of z,
%===
\begin{widetext}
\begin{align}
\begin{aligned}
U(z) &= p^{2} + z \lz \frac{4r_{g}r_{+}\omega}{b}\l r_{+}\omega-\frac{am}{2r_{+}}-\frac{Q^{2}\omega}{2r_{g}}\r-(\Lambda_{lm} + r_{+}^{2}\mu^{2}+a^{2}\omega^{2}) +\frac{q Q }{b}(a m + r_{+} q Q - a^2\omega - 3 r_{+}^{2}\omega)\rz \\
&\quad + z^{2} (a^2 \omega^2 - \Lambda_{lm} + 2\mu^2 a^2 -3 \mu^{2} r_{+}^{2} +6r_{+}^{2}\omega^{2} + 2 Q^2 \mu^2 +  q^2 Q^2  - 6 r_{+} q Q \omega )\\
&\quad + 4z^{3} b \,[r_{g}\mu^{2} + 2r_{+}(\omega^{2}-\mu^{2})-qQ\omega] + 
4z^{4}b^{2}(\omega^{2}-\mu^{2}),\label{eq:U(z)}
\end{aligned}
\end{align}
\end{widetext}
%===
in which,
%===
\begin{align}
\begin{aligned}
    p 
    &= \frac{(r_{+}^{2}+a^{2})}{2b}(\omega-\omega_{c}).
\end{aligned}
\end{align}
%===
Note that both $p$ and $r_g \omega_{c}$ scale as $\mathcal{O}(\alpha^0)$.

In the limit of small $\alpha$, the $\Lambda_{lm}$ has the expanded form $\Lambda_{lm} = l(l+1)+\mathcal{O}(\alpha^4)$. At the LO of $\alpha$, we get the radial equation in limit $(r - r_{+}) \ll \max(1/\omega, 1/\mu)$,
%===
\begin{align}\label{eq:small-r}
z(z+1)\frac{d}{dz}\lz z(z+1)\frac{dR}{dz}\rz+\lz p^{2}-l'(l'+1)z(1+z)\rz R = 0.
\end{align}
%===
At LO, the $l'$ should be replaced by $l$ in this order. Nonetheless, the $\epsilon$ in $l'$ plays the role of a regulator in the intermediate steps. It will be set to zero at the end. 

The general solution of Eq.~\eqref{eq:small-r} is a linear combination of two associated Legendre functions, and the physical solution is the one with the ingoing wave at $r\to r_+$.  After changing the variable back to $r$,  the solution of the radial function is,
%===
\begin{align}
\begin{aligned}
R(r) =\l \frac{r-r_{+}}{r-r_{-}}\r ^{-ip}{_{2}F_{1}}\l -l',l'+1;1-2ip;-\frac{r-r_{+}}{2b}\r,\label{eq:ana-R-small-alpha}
\end{aligned}
\end{align}
%===
up to an arbitrary normalization.

Next, we apply the matching method first proposed in \cite{Detweiler:1980uk} and further developed recently in Ref.~\cite{Bao:2022hew}. The solution of Eq.~\eqref{eq:ana-R-large-r} is only valid in $r\gg r_g$ limit, while the solution in Eq.~\eqref{eq:ana-R-small-alpha} requires $r\ll r_g \alpha^{-2}$ from the ignorance of terms proportional to $z^3$ and $z^4$. They have an overlapped region in the limit $\alpha\ll 1$. In this region, the two solutions are expected to have the same behavior. The behavior of Eq.~\eqref{eq:ana-R-large-r} in the overlapped region is obtained by looking at its small $r$ limit, which is
%============
\begin{align}\label{eq:ana-R-large-r-small-r-limit}
\frac{(2 \kappa )^{l'}\Gamma(-2l'-1)}{\Gamma(-l'-\lambda)} r^{l'}+
\frac{(2\kappa)^{-l'-1} \Gamma(2l'+1)}{\Gamma(l'+1-\lambda)} r^{-l'-1}.
\end{align}
%============
On the other hand, the behavior of Eq.~\eqref{eq:ana-R-small-alpha} in the overlapped region is obtained by looking at its large $r$ limit, which is
\footnote{Without the regulator $\epsilon$, the ratio $\Gamma(-2l-1)/\Gamma(-l)$ in Eq.~\eqref{eq:ana-R-small-r-large-r-limit} is ill-defined and needs to be handled with great caution. In comparison, the calculation with $\epsilon$ is more straightforward. More discussion can be found in the Appendix of Ref.~\cite{Bao:2022hew}.}
%============
\begin{align}\label{eq:ana-R-small-r-large-r-limit}
\begin{split}
\frac{(2b)^{-l'} \Gamma(2l'+1) }{\Gamma(l'+1)\Gamma(l'+1-2ip)} r^{l^\prime} 
+
\frac{(2b)^{l'+1} \Gamma(-2l'-1)}{\Gamma(-l'-2ip) \Gamma(-l')}r^{-l'-1}.
\end{split}
\end{align}
%============
The ratio of the coefficients of the $r^{l'}$ and $r^{-l'-1}$ should be the same for the two solutions in the overlap region. The obtained equation is the eigenequation of $\omega$. It can be solved perturbatively by the observation that the second term in the expression \eqref{eq:ana-R-large-r-small-r-limit} must be suppressed at small $r$, indicating $l'+1-\lambda$ is very close to zero or some negative integer,
%===
\begin{align}\label{eq:n}
l'+1-\lambda = -n-\delta\lambda,
\end{align} 
%===
where $|\delta\lambda|\ll 1$ and $n$ is zero or a positive integer. Following the convention in literature, we also define $\bar{n} = n+l+1$. Then the above relation is re-expressed as $\lambda =\bar{n}+\epsilon+\delta\lambda $. At LO of $\alpha$, it reduces to $\lambda = \bar{n}+\delta\lambda$. Combining with the definition of $\lambda$ in Eq.~\eqref{eq:lambda}, the $r_g\kappa$ scales as $\alpha^2$, which is important in power-counting. Since $|\delta\lambda|\ll 1$, one could solve for $\delta\lambda$ perturbatively with expressions \eqref{eq:ana-R-large-r-small-r-limit} and \eqref{eq:ana-R-small-r-large-r-limit}.

The LO calculation of $\delta\lambda$ for Kerr BHs was completed in Ref.~\cite{Detweiler:1980uk}, with the regulator $\epsilon$ set to zero from the beginning. Recently, we have confirmed a missing factor of $1/2$ in that result \cite{Bao:2022hew}, which was first identified in Ref.~\cite{Pani:2012bp}. The missing factor is conjectured to be from the mistreatment of $\Gamma$ functions with negative integer arguments. The correct formula is provided in the Appendix of Ref.~\cite{Bao:2022hew}. This subtle calculation turns out to be straightforward with the regulator $\epsilon$ kept in the intermediate steps. More details could be found in Ref.~\cite{Bao:2022hew}. For KNBHs, the first LO calculation of $\delta\lambda$ was completed in Ref.~\cite{Huang:2018qdl}. It followed the same steps in Ref.~\cite{Detweiler:1980uk} and missed the factor $1/2$ as well. After the correction, the LO result of $\delta\lambda$ is
%===
\begin{align}\label{eq:DeltaLambda0}
\begin{split}
\delta\lambda^{(0)} =& -ip \,(4\kappa b)^{2l+1}
\frac{(n+2l+1)!(l!)^2}{n!\left[(2l)!(2l+1)!\right]^2} \prod_{j=1}^l (j^2+4p^2),
\end{split}
\end{align}
%===
where the superscript (0) indicates that it is the LO result. It scales as $\mathcal{O}(\alpha^{4l+2})$.

The eigenfrequency $\omega$ can be expressed in terms of $\delta\lambda$ with Eqs.~\eqref{eq:lambda} and~\eqref{eq:n}. Defining $\omega = \omega_{0} + \omega_{1}\delta\lambda^{(0)}$ in Eq.~\eqref{eq:lambda} and expanding it to the linear term of $\delta\lambda^{(0)}$, one arrives at
%===
\begin{align}\label{eq:lambda_expand}
\lambda &= \frac{r_{g}(2\omega_{0}^{2}-\mu^{2})-qQ\omega_{0}}{\sqrt{\mu^{2}-\omega_{0}^{2}}}\nonumber\\
&\quad + \frac{r_{g}\omega_{0}\omega_{1}(3\mu^{2}-2\omega_{0}^{2})-qQ\mu^{2}\omega_{1}}{(\mu^{2}-\omega_{0}^{2})^{3/2}}\delta\lambda^{(0)}
+\mathcal{O}\left((\delta\lambda^{(0)})^2\right).
\end{align}
%===
On the other hand, we have $\lambda = \bar{n}+ \delta\lambda^{(0)}$ from Eq.~\eqref{eq:n}.  Then it is straightforward to get,
%===
\begin{subequations}
\begin{align}
&\frac{r_{g}(2\omega_{0}^{2}-\mu^{2})-qQ\omega_{0}}{\sqrt{\mu^{2}-\omega_{0}^{2}}} = \bar{n}\label{eq:omega0},\\
&\frac{r_{g}\omega_{0}\omega_{1}(3\mu^{2}-2\omega_{0}^{2})-qQ\mu^{2}\omega_{1}}{(\mu^{2}-\omega_{0}^{2})^{3/2}} = 1.\label{eq:omega1}
\end{align}
\end{subequations}
%===
Note that in getting Eq.~\eqref{eq:omega0}, we have ignored the $\epsilon$ which could be traced back to the $l'$ in Eq.~\eqref{eq:n}. This omission leads to an error in $r_g\omega_0$ at the order of $\mathcal{O}(\alpha^5)$. Solving $\omega_0$ perturbatively from Eq.~\eqref{eq:omega0}, one arrives at
%===
\begin{align}
\begin{aligned}
\frac{\omega_{0}^{(0)}}{\mu} = 1-\frac{1}{2} \l\frac{r_{g}\mu-qQ}{\bar{n}}\r^{2} +\mathcal{O}( \alpha^4) \label{eq:omegaapp}.
\end{aligned}
\end{align}
%===
Then the $\omega_{1}$ could be expressed in terms of $\omega_0$ from Eq.~\eqref{eq:omega1} and expanded in powers of $\alpha$,
%===
\begin{align}\label{eq:omega1_omega0}
\begin{aligned}
\frac{\omega_{1}^{(0)}}{\mu} = \frac{(r_g\mu-qQ)^2}{\bar{n}^3} + \mathcal{O}(\alpha^4).
\end{aligned}
\end{align}
%===
Since both $\omega_{0}$ and $\omega_{1}$ are real, $\omega_{0}$ and $\omega_{1}\delta\lambda^{(0)}$ are the leading terms of the real and imaginary parts of $\omega$, respectively. Note, the imaginary part of $\omega$ scales as~$\mathcal{O}(\alpha^{4l+5})$.

\subsection{Next-to-leading-order approximation}\label{sec:AnalyticSolu-NLO}

In a previous work, we have carefully studied the superradiance of a real scalar field around a Kerr BH \cite{Bao:2022hew}. The LO eigenfrequency $\omega$ obtained in Ref.~\cite{Detweiler:1980uk} has an error as large as $160\%$ compared to the numerical result. After correcting the missing factor $1/2$, the convergence is improved, with the error $\lesssim 80\%$. Except for the large discrepancy, the LO result also has some strange behaviors. Since the LO result is the leading term in the Taylor series of the exact $\omega$ at $\alpha=0$, it is expected to converge to the exact $\omega$ with $\alpha$ approaching zero. Nonetheless, the relative error seems to be a nonzero constant for small $\alpha$, reaching as large as $30\%$ at $\alpha =0.07$ for $a=0.99$. This discrepancy at small $\alpha$ calls into question the power-counting strategy. Moreover, the discrepancy at small $\alpha$ increases quickly with the BH spin parameter $a$.

These problems are solved by adding the NLO correction of $\omega$ \cite{Bao:2022hew}. Below we follow the same steps for the KNBHs. The key observation is that the first term in the square bracket in Eq.~\eqref{eq:U(z)}, which scales as $\alpha^2$, is enhanced by a factor of $1/b$. For BHs with large spin $a$ and/or charge $Q$, this term can be as important as the LO contribution. Other NLO contributions are also added for consistency.

The first NLO correction appears as $\epsilon$ in the asymptotic radial wave function at large $r$, which is given in Eq.~\eqref{eq:ana-R-large-r}. It can be calculated from the definition of $l'$ in Eq.~\eqref{eq:lp-def},
%===
\begin{align}
\begin{aligned}
\epsilon = \frac{- 8r_{g}^{2}\mu^{2} + Q^{2}\mu^{2}+8r_{g}qQ\mu-q^{2}Q^{2}}{2l+1}  + \mathcal{O}(\alpha^{4}).
\end{aligned}
\end{align}
%===
The second NLO contribution is from the asymptotic radial wave function at small $r$. The potential $U(z)$ in Eq.~\eqref{eq:U(z)} can be approximated by $p^2-l'(l'+1)z(1+z)+z d$, where $d$ is defined as
%===
\begin{align}
\begin{aligned}
d =& (4r_{g}\mu-2qQ)p-2(4r_{g}-r_{+})r_{g}\mu^{2}\\
&+2\mu qQ(4r_{g}-r_{+})-q^{2}Q^{2} + \mathcal{O}(\alpha^{3}). \nonumber
\end{aligned}
\end{align}
%===
Up to an arbitrary normalization, the corresponding radial function at the NLO is
%=============
\begin{align}\label{eq:ana-R-small-alpha-2}
\begin{split}
R(r) =& \frac{(r-r_-)^{\sqrt{d-p^2}}}{(r-r_+)^{ip}}
\,_2F_1\Big( -l' - ip + \sqrt{d-p^2}, \\
& l'+1 -  ip +\sqrt{d-p^2}; 1 - 2ip;-\frac{r-r_+}{2b}\Big).\\
&~
\end{split}
\end{align}
%=============
In the $r\to +\infty$ limit, the asymptotic behavior of this function is
%=============
\begin{align}
\begin{split}
&\frac{(2b)^{-l'- ip +\sqrt{d-p^2}} \Gamma(2l'+1)\Gamma(1-2ip)}{\Gamma(l'+1-ip-\sqrt{d-p^2})\Gamma(l'+1-ip +\sqrt{d-p^2})} r^{l'}\\
 &+
\frac{(2b)^{l'+1- ip +\sqrt{d-p^2}} \Gamma(-2l'-1)\Gamma(1-2ip)}{\Gamma(-l'-ip-\sqrt{d-p^2})\Gamma(-l'-ip+\sqrt{d-p^2})}r^{-l'-1}.
\end{split}
\end{align}
%=============
Following similar matching steps above, the NLO contribution of $\delta\lambda$ could be obtained after some algebra,
%============
\begin{align}\label{eq:delta-lambda-best}
\delta\lambda^{(1)} = \left(\frac{d}{2\epsilon} -\frac{\epsilon}{2}- ip\right) 
\frac{\left(4\kappa b\right)^{2l'+1}\Gamma(n+2l'+2)\Gamma_{pd}}{n!\left[\Gamma(2l'+1)\Gamma(2l'+2)\right]^2},
\end{align}
%============
where the superscript (1) indicates it is the NLO result, and the $\Gamma_{pd}$ is defined as
%============
\begin{widetext}
\begin{align}
\Gamma_{pd}
=
\frac{\left| \Gamma(l'+1+ip+\sqrt{d-p^2})\Gamma(l'+1+ip-\sqrt{d-p^2})\right|^2 \Gamma(1+2\epsilon) \Gamma(1-2\epsilon)}{\Gamma(1-ip-\sqrt{d-p^2}-\epsilon)\Gamma(1+ip+\sqrt{d-p^2}+\epsilon)
\Gamma(1-ip+\sqrt{d-p^2}-\epsilon)\Gamma(1+ip-\sqrt{d-p^2}+\epsilon)}.
\end{align}
\end{widetext}
%============

The last NLO contribution is from $\omega_0$ and $\omega_1$. Defining $\omega=\omega_0^{(1)}+ \omega_1^{(1)} \delta\lambda^{(1)}$, the expansion of $\lambda$ in Eq.~\eqref{eq:lambda_expand} is still valid, only with $\delta\lambda^{(0)}$ replaced by $\delta\lambda^{(1)}$. Combining with $\lambda = \bar{n} +\epsilon +\delta\lambda^{(1)}$, one could follow the same steps as in the LO calculation and obtain,
%===
\begin{subequations}\label{eq:omega01-NLO}
\begin{align}
\begin{split}
\frac{\omega_0^{(1)}}{\mu} =& 1-\frac{1}{2}\left(\frac{r_g\mu-qQ}{\bar{n}}\right)^2\\
&+\frac{(r_g\mu-qQ)^2}{8\bar{n}^4}\left[3(r_g\mu-qQ)(5r_g\mu-qQ)+8\bar{n}\epsilon\right]\\
&+\mathcal{O}(\alpha^6),
\end{split}\\
\begin{split}
\frac{\omega_{1}^{(1)}}{\mu} =& \frac{(r_g\mu-qQ)^2}{\bar{n}^3} \\
&-\frac{3(r_g\mu-qQ)^2}{2\bar{n}^5} \left[(r_g\mu-qQ)(5r_g\mu-qQ)+2\bar{n}\epsilon\right] \\
&+\mathcal{O}(\alpha^6).
\end{split}
\end{align}
\end{subequations}
%===

Finally, we discuss a subtle problem related to the $\omega$ dependence in the definition of $p$. In the calculation of the $\delta\lambda^{(1)}$, the $\omega$ in $p$ should be replaced by $\omega_0^{(0)}$, rather than $\omega^{(1)}_0$. Here we explain the reason. In deriving the small-$r$ asymptotic form of the radial function, we approximate $U(z)$ in Eq.~\eqref{eq:U(z)} by $p^2-l'(l'+1)z(z+z)+zd$. The coefficient of $z$ and $z^2$ are accurate at $\mathcal{O}(\alpha^2)$ and $\mathcal{O}(\alpha^0)$, respectively. At $z\sim \mathcal{O}(\alpha)$, this two terms are at the same order of $\mathcal{O}(\alpha^4)$. Consequently, we only need to keep the terms in $p^2$ up to $\mathcal{O}(\alpha^4)$, which then leads to $\omega = \omega_0^{(0)}$ in $p$. In comparison to the numerical calculation, this choice of $\omega$ gives a satisfactory NLO result. Using $\omega_0^{(1)}$ in $p$ is not as satisfactory, due to partially including the higher-order contributions.

\section{Results}\label{sec:results}

The eigenfrequency of the Kerr BH superradiance has been studied in Refs.~\cite{Cardoso:2005vk,Dolan:2007mj,Bao:2022hew}. In comparison, the case for Kerr-Newman BH has two more parameters, the BH charge $Q$ and the scalar charge $q$. In this section, we first study the superradiance of a neutral scalar field, focusing on the effect of $Q$. Then we consider the superradiance of a charged scalar field. Comparisons with the numerical calculations in the literature are also provided.

\subsection{Neutral scalar fields}

In the following study of neutral scalar superradiance, we adopt the NLO $\delta \lambda^{(1)}$ in Eq.~\eqref{eq:delta-lambda-best}, where the scalar charge $q$ is set to zero. The $\omega_0^{(1)}$ and $\omega_1^{(1)}$ in Eqs.~\eqref{eq:omega01-NLO} are used. Then the NLO eigenfrequency is $\omega = \omega_0^{(1)}+\omega_1^{(1)} \delta\lambda^{(1)}$.

The BH charge $Q$ cannot be chosen arbitrarily. In our derivation, we have implicitly assumed the KNBH has horizons, which requires $|Q|\leq \sqrt{r_g^2-a^2}$. In addition, neutral scalars could not distinguish the sign of the BH charge. Mathematically, it means the BH charge $Q$ can only appear in the formulas as $Q^2$. So it is sufficient to only consider positive $Q$.

The superradiance condition in Eq.~\eqref{eq:SuperradianceCon} with $q=0$  has the same form as the Kerr BH. The effect of the BH charge $Q$ is hidden in $r_+=r_g+\sqrt{r_g^2-a^2-Q^2}$. Keeping the BH mass $M$ and spin $a$ fixed, larger charge $Q$ results in a larger upper limit of $\text{Re}(\omega)$. Thus massive scalars too heavy to be produced with Kerr BH superradiance may exist in the superradiant region of KNBHs.

%%%%%%%%%%%%%%%%%%%%%%%%%%%%
\begin{figure}[h]
 \begin{center}
 \includegraphics[width=3.3in]{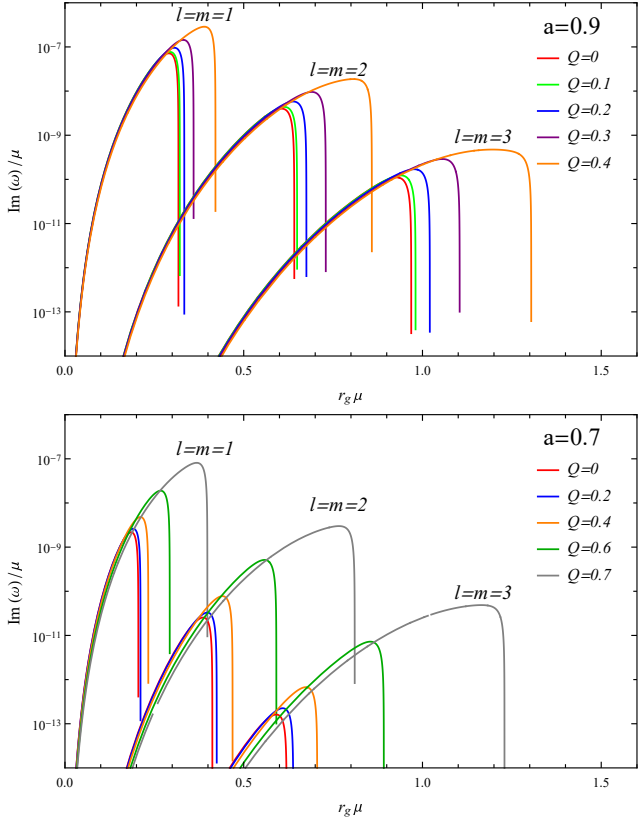}
 \caption{The imaginary part of NLO eigenfrequency with $q=0$ as a function of $r_g\mu$. Only the curves with $n = 0$ are shown. In the top (bottom) panel, the BH spin $a$ is 0.9 (0.7). In both panels, from left to right, the three bunches correspond to $l = m = 1, 2, 3$, respectively. In each bunch, the curves with different colors correspond to different values of the BH charge $Q$.}
 \label{fig:NLOq0}
 \end{center}
\end{figure}
%%%%%%%%%%%%%%%%%%%%%%%%%%%%

%%%%%%%%%%%%%%%%%%%%%%%%%%%%
\begin{figure}[h]
 \begin{center}
 \includegraphics[width=3.3in]{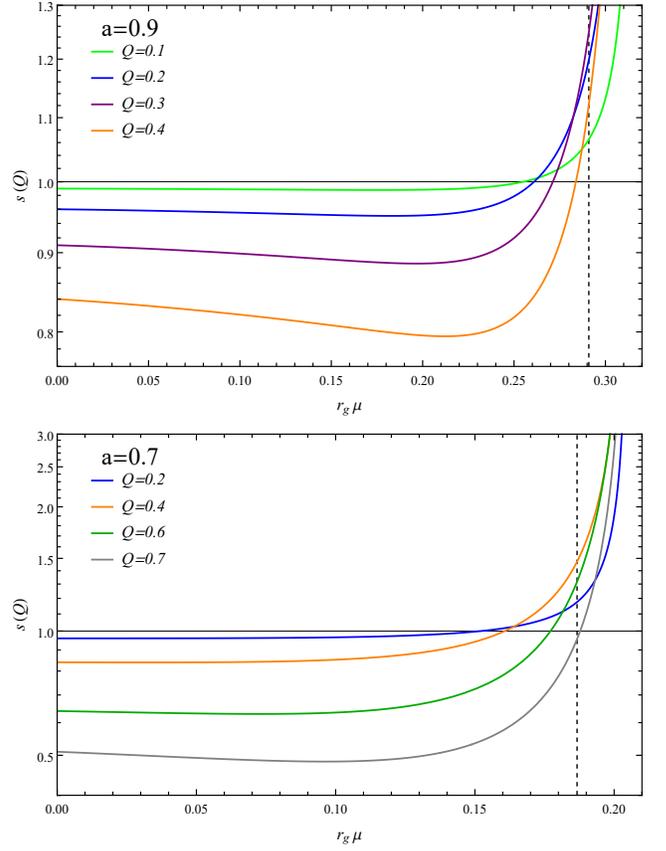}
 \caption{Factor $s(Q)$ with $q=0$ as a function of $r_g\mu$ for BH spin $a=0.9$ (upper panel) and $a=0.7$ (lower panel). The vertical dashed line in each panel labels the value of $r_g\mu$ where  $\text{Im}\,\omega(Q=0)$ reaches its maximum value for the corresponding spin parameter $a$.}
 \label{fig:ratio}
 \end{center}
\end{figure}
%%%%%%%%%%%%%%%%%%%%%%%%%%%%

Figure~\ref{fig:NLOq0} shows the imaginary part of $\omega$ as a function of $r_g\mu$. For comparison, the curves for Kerr BHs are also shown, labeled with $Q=0$. All curves have the same qualitative behavior. With an increasing value of $r_g\mu$, they first increase, then drop rapidly to below zero after reaching the maxima. There are three effects of the BH charge $Q$. Firstly, the superradiant region of $r_g\mu$ is enlarged with larger $Q$. Correspondingly, the peak of the curve moves to the right with increasing $Q$. The maximum $r_g\mu$ with positive $\text{Im}(\omega)$ is quite accurately determined by $\mu=\omega_c$. Secondly, the maximum $\text{Im}(\omega)$ increases with larger $Q$. Fixing the BH spin to be $a=0.9$, the maximum values of $r_g\text{Im}(\omega)$ with $Q=0$ are $2.088\times 10^{-8}$, $2.427 \times 10^{-9}$ and $1.029\times 10^{-10}$ for $l=m=1,2,3$, respectively. The numbers for $Q=0.43$ are $1.476\times 10^{-7}$, $2.006\times 10^{-8}$ and $8.760\times 10^{-10}$, which are larger than the $Q=0$ cases by factors of 7.07, 8.26 and 8.51. For BHs with spin $a=0.7$, the maximum $Q$ is $0.71$. The enhancement factors are 90.29, 269.91, and 707.16, for $l=m=1,2,3$, respectively. Finally, in the ranges of small $r_g\mu$ before reaching the round peaks of the $Q=0$ curves, the charge $Q$ turns out to impede the growth of the scalar clouds. We define a factor $s(Q)$ as
%===
\begin{align}
s(Q) = \frac{\text{Im}\,\omega (Q)}{\text{Im}\,\omega (Q=0)}.
\end{align}
%===
In Fig.~\ref{fig:ratio}, we show $s(Q)$ as a function of $r_g\mu$, for two different BH spins and several values of $Q$. Interestingly, the suppression factor varies slowly with $r_g\mu$. It decreases with increasing $Q$, reaching the minimum value $\sim 0.8$ for $a=0.9$ and $\sim 0.5$ for $a=0.7$.

In Ref.~\cite{Huang:2018qdl}, the authors claim that when $a\gtrsim 0.997 r_g$, the maximum value of $\text{Im}\,\omega$ decreases as $Q$ grows. We do not observe the same behavior. For any spin parameter $a$, the peak value of $\text{Im}\, \omega$ from the NLO approximation increases monotonically with $Q$.

\subsection{Charged scalar fields}

In this part, we study the superradiance of KNBHs under charged scalar perturbation. The NLO eigenfreqency is given by $\omega=\omega_0^{(1)}+\omega_1^{(1)}\delta\lambda^{(1)}$, with the NLO $\delta\lambda^{(1)}$ in Eq.~\eqref{eq:delta-lambda-best}, and the $\omega_0^{(1)}$ and $\omega_1^{(1)}$ in Eqs.~\eqref{eq:omega01-NLO}. Note that the $\omega$ in $p$ should take the form of $\omega_0^{(0)}$ in Eq.~\eqref{eq:omegaapp}, as explained at the end of Sec.~\ref{sec:AnalyticSolu-NLO}. We also compare the NLO results to the LO ones. The latter is given by $\omega=\omega_0^{(0)}+\omega_1^{(0)}\delta\lambda^{(0)}$, with the expressions defined in Eqs.~\eqref{eq:DeltaLambda0}, \eqref{eq:omegaapp} and \eqref{eq:omega1_omega0}. The $\omega$ in $p$  is replaced by $\mu$ for consistency.

%%%%%%%%%%%%%%%%%%%%%
\begin{figure}[h]
 \begin{center}
 \includegraphics[width=3.3in]{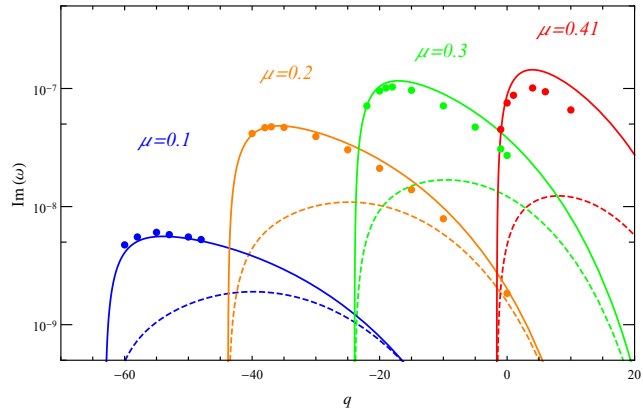}
 \caption{Comparison of the numerical result and the analytic approximations for $n = 0$, $l=m=1$, $a = 0.98$, and $Q = 0.01$, with $r_g$ chosen to be 1 for compacity. The imaginary part of $\omega$ is plotted as a function of the scalar field charge $q$. The dashed (solid) curves are the LO (NLO) approximations and the scattered dots are numerical results taken from Fig.~6 in Ref.~\cite{Furuhashi:2004jk}. The curves with different colors correspond to different values of $\mu$, labeled above the corresponding curves with the same color.}
 \label{fig:Comparison}
 \end{center}
\end{figure}
%%%%%%%%%%%%%%%%%%%%%

Figure~\ref{fig:Comparison} shows the comparison of the LO and NLO approximations to the numerical results taken from Fig.~6 in Ref.~\cite{Furuhashi:2004jk}. The NLO approximation agrees much better with the numerical results. In particular, the average percentage errors of the NLO results for the points in Fig.~\ref{fig:Comparison} are $6.7\%,9.9\%,20.7\%$ and $48.3\%$ for $r_g\mu=0.1, 0.2, 0.3$ and $0.41$, respectively. These numbers can be used as estimates of the NLO results for different values of $\alpha$. Moreover, the convergence of NLO results is better for a smaller value of $r_g\mu$, qualifying the power-counting strategy. To the contrary, the LO results do not seem to converge to the numerical result at small $r_g\mu$, which is also observed for Kerr BHs \cite{Bao:2022hew}. The reason for the bad convergence of the LO result is explained at the beginning of Sec.~\ref{sec:AnalyticSolu-NLO}. A caveat is that the curves for the LO approximations in Fig.~\ref{fig:Comparison} are not the same as those in Ref.~\cite{Furuhashi:2004jk}. The latter misses a factor of $1/2$.

Table.~\ref{tab:Numerical} shows the comparison of the NLO results and the numerical solutions for five more parameter sets in the literature. They are the most unstable modes with different parameters. The percentage uncertainty of the NLO approximation varies from 14\% to 29\% compared to the numerical results.

%%%%%%%%%%%%%%%%%%%%%%%
\begin{table}[h]
\caption{\label{tab:Numerical}
Comparison of the NLO approximations of Im($\omega$) with the numerical results from Ref.~\cite{Huang:2018qdl} (cases A to D) and from Ref.~\cite{Furuhashi:2004jk} (case E). All cases are with $n=0$ and $l = m = 1$. The numbers below assume $r_g=1$ for compacity. The percentage error is calculated by taking the difference between the approximation and the numerical result, then dividing it by the numerical result.\\
Case A: $a=0.9$, $Q=0.2$ , $q=-0.264$, $\mu=0.282$; \\
Case B: $a=0.99$, $Q=0.1105$ , $q=-0.6335$, $\mu=0.397$; \\
Case C: $a=0.997$, $Q=0.004$ , $q=-18.91$, $\mu=0.39822$; \\
Case D: $a=0.997$, $Q=0.0001$, $q=-756.68$, $\mu=0.39816$;\\ 
Case E: $a=0.98$, $Q=0.01$, $q=-8$, $\mu=0.35$. }
\begin{ruledtabular}
\begin{tabular}{cccc}
Case & Type &Im($\omega$) & $\%$ error  \\
\hline
&LO & 5.623$\times 10 ^{-9}$ & 74.9$\%$\\
A&NLO & 2.882$\times 10 ^{-8}$ & 28.5$\%$\\
&Numerical& 2.243$\times 10 ^{-8}$& - \\
\hline
&LO & 1.224$\times 10 ^{-8}$& 92.9$\%$\\
B&NLO  & 1.981$\times 10 ^{-7}$& 14.1$\%$\\
&Numerical& 1.736$\times 10 ^{-7}$& -\\
\hline
&LO& 1.264$\times 10 ^{-8}$ & 92.9$\%$\\
C&NLO & 2.041$\times 10 ^{-7}$ & 14.1$\%$\\
&Numerical& 1.788$\times 10 ^{-7}$ &-\\
\hline
&LO & 1.263$\times 10 ^{-8}$& 92.9$\%$\\
D&NLO & 2.041$\times 10 ^{-7}$& 14.1$\%$\\
&Numerical& 1.788$\times 10 ^{-7}$& -\\
\hline
&LO& 1.27$\times 10 ^{-8}$ & 88.8$\%$\\
E&NLO & 1.39$\times 10 ^{-7}$ & 22.7$\%$\\
&Numerical& 1.13$\times 10 ^{-7}$ & -\\
\end{tabular}
\end{ruledtabular}
\end{table}
%%%%%%%%%%%%%%%%%%%%%%%

Next, we analyze the effect of $q$. In the formulas, the $q$ and $Q$ appears as $qQ$ and $Q^2$. So it is sufficient to consider the case with $Q>0$, and with $q$ being either positive or negative. There are two constraints for the existence of the superradiant bound states. The superradiance requires $\omega<\omega_c$ in Eq.~\eqref{eq:SuperradianceCon}. And the existence of the bound states gives the second constraint $\lambda>0$ from Eq.~\eqref{eq:lambda}, which is approximately $r_g\mu-qQ>0$.

%%%%%%%%%%%%%%%%%%%%%%%%%%%
\begin{figure}[h]
\begin{center}
\includegraphics[width=3.5in]{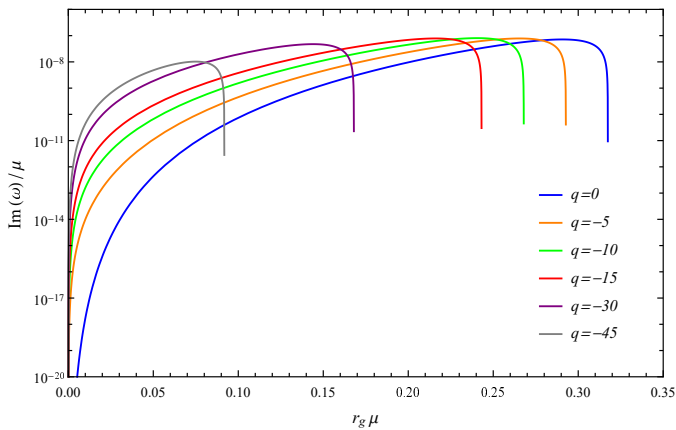}
\caption{The imaginary part of NLO eigenfrequency as a function of $r_g\mu$ with different negative values of $q$. Other parameters are $n = 0$, $l=m=1$, $a=0.9$ and $Q=0.01$.}
\label{fig:qL0}
\end{center}
\end{figure}
%%%%%%%%%%%%%%%%%%%%%%%%%%%

%%%%%%%%%%%%%%%%%%%%%%%
\begin{table}
\caption{\label{tab:Max}
The maximum value of $\text{Im}(\omega)$ obtained by varying $q$, with $a$ and $Q$ fixed. The numbers below assume $r_g=1$ for compacity. }
\begin{ruledtabular}
\begin{tabular}{ccc}
(a,Q) & q &Im($\omega$)  \\
\hline
&-2.5 & 2.10313$\times 10 ^{-8}$\\
(0.9, 0.01)&-2.25& 2.10329$\times 10 ^{-8}$\\
&-2.2& 2.10329$\times 10 ^{-8}$\\
&-2& 2.10268$\times 10 ^{-8}$\\
\hline
&-1.25 & 2.10814$\times 10 ^{-8}$\\
(0.9, 0.02)&-1.1 & 2.10831$\times 10 ^{-8}$\\
&-1& 2.10815$\times 10 ^{-8}$\\
&-0.75& 2.10682$\times 10 ^{-8}$\\
\hline
&-3& 4.14247$\times 10 ^{-10}$ \\
(0.7, 0.01)&-2.8 & 4.14270$\times 10 ^{-10}$ \\
&-2.75& 4.14260$\times 10 ^{-10}$ \\
&-2.5& 4.14104$\times 10 ^{-10}$ \\
\hline
&-1.5& 4.14863$\times 10 ^{-10}$ \\
(0.7, 0.02)&-1.4 & 4.14888$\times 10 ^{-10}$ \\
&-1.25& 4.14726$\times 10 ^{-10}$ \\
&-1& 4.13927$\times 10 ^{-10}$ \\
\end{tabular}
\end{ruledtabular}
\end{table}
%%%%%%%%%%%%%%%%%%%%%%%

If the scalar and the KNBH at the center have opposite charges, i.e. $qQ<0$, the scalar cloud is more tightly bounded. In this case, the second constraint above is automatically satisfied. Figure~\ref{fig:qL0} shows the imaginary part of $\omega$ as a function of $r_g\mu$ in the $n=0$, $l=m=1$ bound state, with BH spin $a=0.9$ and charge $Q=0.01$. The scalar charge $q$ varies from $-45$ to $0$. The region of superradiance shrinks when $q$ is more negative, which is a consequence that $\omega_c$ decreases with $q$ for fixed $Q$. The peak value of $\text{Im}(\omega)$ seems to be smaller with decreasing $q$. Nonetheless, a more careful study shows that the maximum $\text{Im}(\omega)$ happens at some small but nonzero $|q|$ (see Table.~\ref{tab:Max}).

%%%%%%%%%%%%%%%%%%%%%%%%%%%
\begin{figure}[h]
\begin{center}
\includegraphics[width=3.5in]{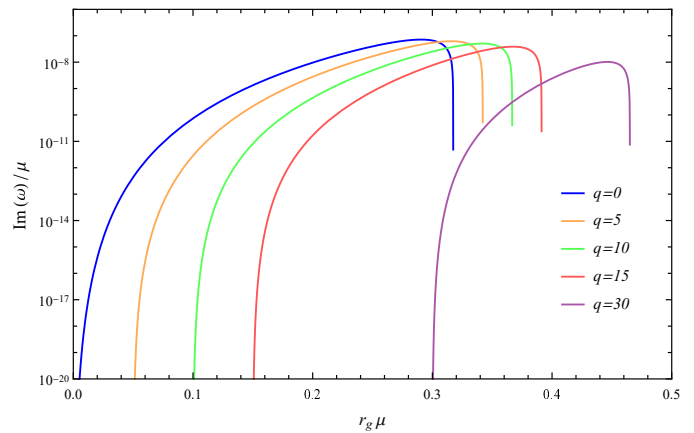}
\caption{
The imaginary part of NLO eigenfrequency as a function of $r_g\mu$ with different positive values of $q$. Other parameters are $n = 0$, $l=m=1$, $a=0.9$ and $Q=0.01$.}
\label{fig:qG0}
\end{center}
\end{figure}
%%%%%%%%%%%%%%%%%%%%%%%%%%%

If the charges of the scalar and the KNBH have the same sign, i.e. $qQ>0$, the scalar cloud is less bounded. The second constraint above gives $r_g\mu> qQ$ for the existence of bound states. Figure~\ref{fig:qG0} shows the imaginary part of $\omega$ as a function of $r_g\mu$ in the $n=0$, $l=m=1$ bound state, with BH spin $a=0.9$ and charge $Q=0.01$. With larger value of positive $q$, the superradiance region shrinks and the peak is lower as well.

\section{Conclusion}\label{sec:Conclusion}

In this work, we have studied the scalar superradiant instability of the KNBH and obtained the LO and NLO expressions of the superradiant rate in the regime of $\alpha\ll~\!\!\!1$. The calculation is based on the matching method which is proposed by Detweiler for Kerr BHs in Ref.~\cite{Detweiler:1980uk} and developed in our previous work \cite{Bao:2022hew}. In this paper, we further refine the power-counting strategy and apply it to the KNBH.

The LO scalar superradiant rate for KNBH has been calculated previously in Ref.~\cite{Furuhashi:2004jk}. With our refined power-counting strategy, a similar result is obtained but with an extra overall factor of $1/2$. We conjecture the factor is from the mistreatment of the $\Gamma$ functions with negative integer arguments, similar to the case of Kerr BHs. More analysis could be found in our previous work \cite{Bao:2022hew}.

We compare the LO and NLO results with the existing numerical calculations in the literature. The LO results are smaller than the numerical solutions by an order of magnitude. To the contrary, the percentage error of the NLO result ranges from a few percent to about $50\%$, depending on the value of $\alpha$ (see Fig.~\ref{fig:Comparison} and Table~\ref{tab:Numerical}). In particular, the error of the NLO result decreases for a smaller value of $\alpha$, qualifying our power-counting strategy.

The obtained NLO expression has a compact form and can be straightforwardly applied to phenomenological studies of the KNBH superradiance as well as the ultralight scalars, either neutral or charged. Besides the superradiance condition $\text{Re}(\omega) < m\Omega_H$ as the Kerr BHs, there is another condition $r_g\mu>qQ$ for the existence of bound states. For neutral scalars, larger BH charge $Q$ leads to a larger superradiant range of $r_g\mu$ as well as the maximum superradiant rate (see Fig.~\ref{fig:NLOq0}). Thus massive neutral scalars too heavy to be produced with Kerr BH superradiance may exist in the superradiant region of KNBHs. The situation is different for charged scalars. For fixed BH spin $a$ and charge $Q$, increasing the scalar charge $q$ always leads to narrower superradiant range of $r_g\mu$ (see Figs.~\ref{fig:qL0} and~\ref{fig:qG0}). Interestingly, the maximum superradiant rate happens at a small negative scalar charge $q$ (see Table~\ref{tab:Max}). We have no explanation for this observation.

\begin{acknowledgments}
This work is supported in part by the National Natural Science Foundation of China (NSFC) under Grant No. 12075136 and the Natural Science Foundation of Shandong Province under Grant No. ZR2020MA094.
\end{acknowledgments}

%%%%%%%%%%%%%%%%%%%%%%%%%%%%%%%%%%%%%%%%%%
%%%%%%%%%% Bibliography %%%%%%%%%%%%%%%%%%
%%%%%%%%%%%%%%%%%%%%%%%%%%%%%%%%%%%%%%%%%%

\end{document}